\documentclass[12pt,preprint]{aastex}

\renewcommand {\deg}   {\mbox{$^\circ$}}

\newcommand   {\arcs}  {\mbox{$^{\prime\prime}$}}

\newcommand   {\kms}   {\mbox{km\,s$^{-1}$}}
\renewcommand {\ga}    {\mbox{\rlap{\hbox{\lower5pt\hbox{$\sim$}}}\hbox{$>$}}}
\renewcommand {\la}    {\mbox{\rlap{\hbox{\lower5pt\hbox{$\sim$}}}\hbox{$<$}}}

\begin{document}



\def\kms {\hbox{km{\hskip0.1em}s$^{-1}$}} 
\def\msol{\hbox{$\hbox{M}_\odot$}}
\def\lsol{\hbox{$\hbox{L}_\odot$}}
\def\kms{km s$^{-1}$}
\def\Blos{B$_{\rm los}$}
\def\etal   {{\it et al. }}                     
\def\psec           {$.\negthinspace^{s}$}
\def\pasec          {$.\negthinspace^{\prime\prime}$}
\def\pdeg           {$.\kern-.25em ^{^\circ}$}
\def\degree{\ifmmode{^\circ} \else{$^\circ$}\fi}
\def\ee #1 {\times 10^{#1}}          
\def\ut #1 #2 { \, \textrm{#1}^{#2}} 
\def\u #1 { \, \textrm{#1}}          
\def\nH {n_\mathrm{H}}

\def\ddeg   {\hbox{$.\!\!^\circ$}}              
\def\deg    {$^{\circ}$}                        
\def\le     {$\leq$}                            
\def\sec    {$^{\rm s}$}                        
\def\msol   {\hbox{$M_\odot$}}                  
\def\i      {\hbox{\it I}}                      
\def\v      {\hbox{\it V}}                      
\def\dasec  {\hbox{$.\!\!^{\prime\prime}$}}     
\def\asec   {$^{\prime\prime}$}                 
\def\dasec  {\hbox{$.\!\!^{\prime\prime}$}}     
\def\dsec   {\hbox{$.\!\!^{\rm s}$}}            
\def\min    {$^{\rm m}$}                        
\def\hour   {$^{\rm h}$}                        
\def\amin   {$^{\prime}$}                       
\def\lsol{\, \hbox{$\hbox{L}_\odot$}}
\def\sec    {$^{\rm s}$}                        
\def\etal   {{\it et al. }}                     

\def\xbar   {\hbox{$\overline{\rm x}$}}         

\slugcomment{Submitted to ApJL}
\shorttitle{}
\shortauthors{}

\title{Cosmic-Ray Heating of Molecular Gas in the Nuclear
Disk: Low  Star Formation Efficiency}
\author{F. Yusef-Zadeh\altaffilmark{1},
M. Wardle\altaffilmark{2},
S. Roy\altaffilmark{3}}

\altaffiltext{1}{Department of Physics and Astronomy,
Northwestern University, Evanston, Il. 60208
(zadeh@northwestern.edu)}
\altaffiltext{2}{Department of Physics, Macquarie University, Sydney NSW 2109,
Australia (wardle@physics.mq.edu.au)}
\altaffiltext{3}{Astron, P.O. Box 2, 7990 AA, Dwingeloo, The
Netherlands (roy@astron.nl)}

\begin{abstract} Understanding the processes occurring in the nuclear
disk of our Galaxy is interesting in its own right, as part of the
Milky Way Galaxy, but also because it is the closest galactic nucleus.
It has been more than two decades since it was recognized that the
general phenomenon of higher gas temperature in the inner few hundred
parsecs by comparison with local clouds in the disk of the Galaxy.
This is one of the least understood characteristics of giant molecular
clouds having a much higher gas temperature than dust temperature in
the inner few degrees of the Galactic center.  We propose that an
enhanced flux of cosmic-ray electrons, as evidenced recently by a
number of studies, are responsible for directly heating the gas clouds
in the nuclear disk, elevating the temperature of molecular gas
($\sim$ 75K) above the dust temperature ($\sim$ 20K).  In addition we
report the detection of nonthermal radio emission from Sgr B2-F based
on low-frequency GMRT and VLA observations.  The higher ionization
fraction and thermal energy due to the impact of nonthermal electrons
in star forming sites have important implications in slowing down star
formation in the nuclear disk of our galaxy and nuclei of galaxies.

\end{abstract}

\section{Introduction}
\label{introduction} 

The nuclear disk of our Galaxy has been studied extensively in
molecular lines at millimeter wavelengths.  Multi-transition ammonia
observations have probed the temperature of gas (G\"usten, Walmsley
and Pauls 1981; Morris et al.  1983; G\"usten et a.  1985;
H\"uttemeister et al.  1993), which was measured to be in the range of
50--120 K and was found to be uniformly high throughout the inner 500
pcs of the Galaxy.  A high spatial resolution study of 36 clouds
between l=$-1^0$ and 3$^0$ (H\"uttemeister et al.  1993) found a
two-temperature distribution, warm low-density gas
(T$_{kin}\sim200$\,K, $n(\mathrm{H}_2)\sim10^3$\,cm$^{-3}$) and oool
dense cores (T$_{kin}\sim25$K, $n(\mathrm{H}_2)\sim10^5$\,cm$^{-3}$).
In another study, ISO observations of rotational transitions of H$_2$
found predominantly warm molecular gas with $T \sim 150$\,K toward 16
Galactic center molecular clouds with H$_2$ column densities of
$\sim1-2\times10^{22}$\,cm$^{-2}$ (Rodriguez-Fern\'andez et al.
2001).  A recent absorption line study of the metastable (3,3)
rotational level of H$_3^+$ suggests that there is a large quantity of
warm (T$\sim$250 K) and diffuse ($n(\mathrm{H}_2)\sim100$\,cm$^{-3}$)
gas distributed in the central few hundred pcs of the Galaxy (Oka et
al.\ 2005).

Molecular gas with $T \ga 100$\,K in the Galactic disk is heated by
collisions with grains that have been warmed by hot stars in star
forming regions.  Thus, regions of high kinetic temperatures inferred
in NH$_3$ observations of star forming regions is strongly correlated
with high dust temperature of clouds that accompany IR sources.
However, the high gas temperature in GMCs in the nuclear disk (Lis et
al.  2001) is inconsistent with the dust temperature 18-22\,K inferred
toward the inner 2$^0\times1^0$ of the Galaxy from SCUBA 850 and 450
$\mu$m and IRAS observations (Cox \& Laurejs 1989; Pierce-Price et al.
2000).

A global heating mechanism is needed to explain the significantly
higher gas temperature than dust temperature in a large fraction of
gas and dust clouds in the nuclear disk.  With the exception of Sgr B2,
the lack of embedded sources able to provide significant heating is
supported by the paucity of 6.7 MHz methanol sources in this region
(Caswell 1996).  High ionization by a large flux of cosmic rays
heating Galactic center molecular clouds has been suggested by a
number of authors (G\"usten et al.\ 1981; H\"uttemeister et al.
1993).  Alternative suggestions such as cloud-collisions or global
fast shocks have also been made but there is no clear observational
evidence to support these hypotheses (Martin-Pintado et al.\ 1997; Lis
et al.  2001).  Here, we describe several recent studies indicating an
excess cosmic ray flux in the central region of the Galaxy and revisit
the cosmic ray heating scenario initially proposed by G\"usten et al.
(1981) .  We also report the detection of nonthermal radio emission
from the massive star forming region Sgr~B2 followed by the
implications of cosmic ray interaction with molecular clouds.  In
particular, the consequences of such interaction is discussed in the
context of star formation in the Galactic center region. 

\section{Evidence for Enhanced Cosmic Ray Flux}

Our interest in reconsidering the global heating of molecular clouds
by cosmic rays in the central 500 pcs stem from three different
studies which infer enhanced cosmic rays there.  
First, the fluorescent 6.4 keV K$\alpha$ iron line emission from the
G0.11--0.08 molecular cloud, which lies $\sim$15 pcs in projection
from the Galactic center (Tsuboi et al.\ 1997; Oka et al.\ 2001), is
accounted for by he impact of low-energy cosmic ray electrons
(Yusef-Zadeh, Law \& Wardle 2002).  The energy density over the 10$^6$
\msol\ cloud was estimated to be 1-2 eV cm$^{-3}$, increasing to 150
eV cm$^{-3}$ at the edge of the cloud where there it interacts with a
nonthermal radio filament.  
The required energy density of cosmic rays at the edge of
the cloud increases to  $\sim$250 eV cm$^{-3}$ if the 
the molecular mass of  G0.11-0.08 is 6$\times10^5$ \msol (Handa et al.  2006). 
  This idea was also applied to other
prominent Galactic center molecular clouds from which diffuse 6.4 keV
line emission is detected (Yusef-Zadeh et al.\ 2007a).  The energy
density of the cosmic rays required to explain the observed X-ray
emission from the clouds in the inner 2$^0\times 0.5^0$ of the Galaxy
ranges between 20 and 10$^3$ eV cm$^{-3}$.  A cosmic-ray energy
density of 0.2 eV cm$^{-3}$ is required to explain the Galactic
ridge X-ray emission (Valinia et al.\ 2000).  The inferred ionization
rates of the Galactic center clouds based on the 6.4 keV line
measurements range between 2$\times10^{-14}$ and 5$\times10^{-13}$
s$^{-1}$ H$^{-1}$ (Yusef-Zadeh et al.\ 2002, 2007a).

Second, strong H$_3^+$ absorption along several lines of sight towards
the Galactic center has been reported by Oka et al.\ (2005), who
inferred that the the ionization rate in this unique environment
ranges between 2--7 $\times10^{-15}$ s$^{-1}$.  Furthermore, an
H$_3$O$^+$ study toward Sgr B2, one of the densest clouds in the
Galaxy, inferred also an ionization rate of $\sim4\times10^{-16}$
s$^{-1}$ (van der Tak et al.\ 2006).

Third, the detection of low frequency 74 MHz radio emission from the
central disk of the Galaxy indicates enhanced cosmic rays from the
central degree of the Galaxy.  LaRosa et al.\ (2005) estimate that the
cosmic ray electron density of the central 1.5$\times0.5$ degrees is
$\sim$7.2 eV cm$^{-3}$, about 15 times higher than that in the local
ISM (Webber 1998).  In addition, detailed spectral index measurements
of extended radio sources show that 85$\pm4$\% of 6\,cm continuum
emission corresponding to a flux density of 841$\pm$44 Jy from the
inner $\sim2.5^0\times1^0$ of the Galaxy is nonthermal (Law 2007).

\section{GMRT and VLA Radio Continuum Analysis}

Motivated by the strong detection of 6.4 keV line emission from Sgr
B2, we searched for evidence of nonthermal continuum emission from
this cloud.  To examine the picture of cosmic-rays impacting molecular
gas in Sgr B2, observations were conducted on March 14, 2003 using the
Giant Meterwave Radio Telescope (GMRT) at 255 and 583 MHz with an
effective bandwidth of 6 MHz using the default spectral line mode of
the correlator.  The field center was set at $\alpha, \delta
(J2000)=17^h46^m00^s, -28d57'00^{"}$.  3C48 was used as primary flux
density calibrator, and 1830-36 was used as secondary calibrator.  The
data were processed using standard programs in AIPS. After calibration
and editing of the 255 MHz data, a pseudo-continuum database of 9
frequency channels was made from the central 5.6 MHz of the observed 6
MHz band.  Images of the fields were formed after the application of
phase self-calibration.

In addition, we used archival data taken with the Very Large Array
(VLA) of the National Radio Astronomy Observatory\footnote{The
National Radio Astronomy Observatory is a facility of the National
Science Foundation, operated under a cooperative agreement by
Associated Universities, Inc.} at 1.4 GHz and 327 MHz.  The data
reduction is described in Yusef-Zadeh, Hewitt, \& Cotton (2004)
and Nord et al.\ (2004), respectively.

Figure 1a shows contours of 255 MHz radio continuum emission
superimposed on a grayscale image of Sgr B2 at 327 MHz.  The peak
emission at 327 MHz centered at $\alpha, \delta(J2000)=17^h47^m201.6^s, 
-28^023'03.63''$ coincides with the position of the brightest cluster
of 20 ultracompact HII regions known as the F cluster in Sgr B2 Main
(DePree et al.\ 1998), To determine the flux density of this source at
1.4 GHz and 255 MHz, we used Gaussian-fitted fluxes from background
subtracted images.  Figure 1b shows the spectrum of the emission based
on images convolved to $22.5''\times16.8''$ resolution.  The high
frequency emission from Sgr B2 is due to bright, compact and optically
thick HII regions whose flux densities should drop with decreasing
frequencies.  Increased free-free absorption by foreground material is
responsible for the drop in the flux density at low frequencies (see
below).
 
Despite the complexity of bright radio emission from Sgr B2, there is
evidence for nonthermal emission at low frequencies.  
First, it is clear that the isolated emission from the F cluster
dominates this complex region at low frequencies but has similar
surface brightness to Sgr B2 North at high frequencies.  The 1490 MHz
images of the Sgr B2 M and N show peak flux densities of $\sim$1.1 Jy
within a beam size of 10$''\times10''$ (Yusef-Zadeh, Hewitt and Cotton
2004).

Second, the spectral index $\alpha$, where F$_{\nu}\propto
\nu^{\alpha}$, between 255 and 327 MHz is estimated to be 1.28$\pm0.4$
whereas $\alpha$ is 2.35$\pm0.30$ between 1490 and 583 MHz.  A flatter
spectral index suggests a steep spectrum emission is contributing to
the flux of Sgr B2 at low frequencies.  This inference is strengthened
by the fact that the measured flux density at 255 MHz would be higher
had the effect of foreground free-free absorption been accounted for.
Assuming that the free-free absorption optical depth is $\sim 1$ at
150 MHz (Roy \& Rao 2006), is $\propto \lambda^{2.1}$, and is the same
toward both Sgr B2 and the Sgr A complex, then $\tau\sim0.33$ and 0.19
at 255 and 327 MHz, respectively.  The unabsorbed flux density of Sgr
B2 region would then imply $\alpha=0.81\pm0.4$ between 255 and 327
MHz.  This shows that the radio spectrum becomes flatter at lower
frequencies, as shown in Figure 1b.  Furthermore, if we extrapolate
the flux density at 1490 MHz to 255 MHz with a $\sim$2.3 spectral
index, the flux density is expected to be $\sim160$ mJy which is lower
than the unabsorbed flux density of 242 mJy at 255 MHz.  The
difference between the measured and estimated flux densities at 255
MHz is 82 mJy.  Assuming that the thermal and nonthermal components
are distinct, the nonthermal component at 255 MHz is estimated to be
82 mJy.

Third, two studies of nonthermal emission from Sgr B2 provide further
evidence for cosmic rays associated with Sgr B2 (Hollis et al.\ 2007;
Crocker et al.\ 2007).  Detailed GBT observations of Sgr B2 report a
flux $\sim$4.6 Jy at 44 GHz with $\alpha$=-0.7 (Hollis et al.\ 2007).
The nonthermal radio flux needed 
to explain the  6.4 keV line
emission from Sgr B2  is 18 Jy at 327 MHz 
(Yusef-Zadeh et al.\ 2007a).  
Thus, the observed  nonthermal flux measured with GBT 
is more than enough to account for  the origin of the observed 6.4 keV line
emission from Sgr B2.  We believe a fraction of the
observed nonthermal emission from Sgr B2 is localized with the Sgr B2
F cluster and the rest is diffuse over a larger extent associated with
Sgr B2.
 
Finally, radio continuum and radio recombination line studies indicate
an unusually high electron temperature, high brightness temperature,
rising spectral index and an anomalously low helium abundance at the
position of the F cluster in Sgr B2 (Mehringer et al.\ 1995; DePree et
al.  1998).  A nonthermal contribution to the continuum emission could
be responsible for the low line-to-continuum ratio, and
thus the inferred high brightness temperature 
as well as the abundance of atomic and molecular species 
affected  by the bombardment of cosmic ray particles.  A more
detailed account will be given elsewhere.

To estimate the energy density of electrons we adopt a typical
nonthermal flux of 80\,mJy at 255\,MHz in a $22.''5 \times 16.''8$
beam, with a $\nu^{-0.7}$ spectrum.  The relativistic electrons
responsible for the emission must have an $E^{-2.4}$ energy spectrum,
and we suppose that this extends down to 1\,MeV. Their energy density
is $\sim 3 \u eV \ut cm -3 $ for $B=1$\,mG or $\sim 150 \u eV \ut cm
-3 $ for $B=0.1$\,mG. The corresponding ionization rate can be
estimated by noting that in the MeV range the stopping power of the
ISM is about 3.5 $\u MeV \ut g -1 \ut cm 2 $ (ICRU 1984), and that on
average one ionization occurs for each 40.1\,eV deposited into the gas
(Dalgarno, Yan \& Liu 1999).  This yields an ionization rate $\sim
2\ee -14 \ut s -1 \ut H -1 $.  The energy density and ionization rate
are reduced by a factor $\sim 25$ if the electron spectrum only
extends down to 10\,MeV instead of 1\,MeV.

\newcommand\nh{n_\mathrm{H}}
\newcommand\percc{\textrm{cm}^{-3}}
\section{Discussion}

\subsection{Effects of Enhanced Cosmic Ray Fluxes in Star Forming Regions}

High cosmic-ray fluxes in molecular clouds affect star formation by
heating the gas and increasing its ionization fraction.  Higher cloud
temperatures increase the Jeans mass, potentially changing the IMF,
while high ionization increases magnetic coupling to the cloud
material, reducing ambipolar diffusion and increasing the time
taken for gravitationally unstable cores to contract to the point that
they overwhelm their magnetic support.  

The heating associated with cosmic ray ionizations can be estimated as
follows.  Each ionization of a hydrogen molecule is associated on
average with 40.1 eV energy loss by electrons, of which 11\% ends up
as heat (e.g.\ Dalgarno et al.\ 1999).  In addition another 8\,eV  
appears as heat when H$_3^+$ recombines (e.g.\ Maloney, Hollenbach \& Tielens
1996). Thus,  each ionization of a
hydrogen molecule is associated with the deposition of 12.4\,eV of heat
into the gas. As the ionization rate per hydrogen \emph{nucleus} is
half the H$_2$ ionization rate $\zeta_\mathrm{H}$, the heating rate
per hydrogen nucleus $\Gamma/\nh$ is $\approx$ 25\,eV$\times \zeta_\mathrm{H}$, or
\begin{equation}
    \Gamma/\nh = 4.0\ee -26 \left(\frac{\zeta_\mathrm{H}}{10^{-15}\ut s -1
    \ut H -1 }\right) \u erg \ut s -1 \ut H -1
    \label{eq:Gamma}
\end{equation}
As pointed out by G\"usten et al.\ (1981), an ionization rate $\zeta_H
\sim 1\ee -15 \ut s -1 $ is sufficient to explain the observed gas
temperature of $\sim 70$\,K. Using the cooling rates calculated by
Neufeld, Lepp \& Melnick (1995) for $n(\mathrm{H}_2) = 5000\ut cm -3 $
and $N(\mathrm{H}_2)/\Delta v = 10^{22} \ut cm -2 \ut km -1 \u s $, the
equilibrium temperatures are approximately 60, 130 and 280\,K for
$\zeta_H = 10^{-15}$, $10^{-14}$ and $10^{-13}\ut s -1 \ut H -1 $
respectively.

The Jeans mass can be estimated by equating the free-fall time of a uniform 
cloud core of density $\rho$ and radius $R$, i.e.\ $t_{ff} = 1/\sqrt{G
\rho}$, to the sound
crossing time $R/c_s$, yielding
\begin{equation}
    M_J \approx 11 \left(\frac{T}{75\u K }\right)^{3/2}
    \left(\frac{\nH}{10^6 \ut cm -3 }\right)^{-1/2}  \msol
    \label{eq:M_J}
\end{equation}

Collapse of this Jeans-unstable core is halted by the cloud's magnetic
field if the mass-to-flux ratio is less than the critical value
$1/\sqrt{4\pi G}$, ie.  if $B\,\ga\, 0.1 \nh\,/\,(10^6\percc) \u mG $.
However the magnetic support is temporary because the cloud is
weakly-ionized, and the predominant neutral species are able to drift
towards the centre under the action of gravity while colliding with
the ions and electrons that are tied to and supported by the
near-static field lines.  The neutral drift speed is determined by the balance
between gravity and the drag due to collisions with the ions:
\begin{equation}
    \frac{GM\rho}{R^2} \approx n_i <\sigma v> \rho v_d
    \label{eq:AD_forces}
\end{equation}
where $<\sigma v> \approx 2\ee -9 \ut cm -3 \ut s -1 $ is the rate
coefficient for ion-neutral momentum transfer.  This yields
a drift speed of a few hundredths of a kilometer per second.
This drift increases the mass-to-flux ratio at the core's centre on a time scale
\begin{equation}
    t_\mathrm{AD} = \frac{R}{v_d} \approx 0.8 \left(\frac{x_e}{10^{-8}}\right) 
    \u Myr
    \label{eq:tad}
\end{equation} 
until it attains the critical value at which point dynamical collapse
occurs on a few free-fall times.  A more accurate calculation by
Mouschovias (1987) reduces this estimate by a factor of two.  The
ambipolar diffusion timescale is of order a few Myr for the standard
interstellar ionization rates but is directly proportional to the
ionization fraction, or equivalently to the square root of the
ionization rate.  Thus the time scale to achieve a supercritical core
becomes large if the ionization rate is increased a hundred-fold
over standard values.\footnote{see Le Petit et al.\ (2004) and
references therein for recent determinations of the ionization rate
in diffuse clouds}

Recent observations indicate that massive star formation has
signatures similar to those seen in low-mass star formation: low star
formation efficiency (Krumholz \& Tan 2007), disc accretion (e.g.
Cesaroni et al.\ 2005) and molecular outflows (e.g. Zhang et al.
2005).  If the initial phases of high-mass star formation are
analogous to low-mass star formation, then the formation of high mass
stars should be affected by the increased cosmic-ray ionization rate.
Due to the higher Jeans mass in the warm gas, more massive stars are
expected to preferentially form in the nuclear disk.  This scenario is
consistent with recent observations of a number of unique and young
massive stellar clusters with a top heavy IMF (Figer et al.\ 2004;
Stolte et al.\ 2005; Nayakshin \& Sunyaev 2005).

In this environment, the strong tidal shear will allow only the
densest clouds to survive.  Survival against shear requires that the
gravitational frequency of the cloud, $\sqrt{G\rho}$, must exceed the
orbital frequency around the Galactic Centre, $v/R$.  Adopting a
galctocentric distance $R=100$ pc and an orbital speed $v=150$\,km/s,
we find that clouds are tidally stable only if $\nH > 1.5\times 10^4
\percc $.

\subsection{Implications \& Conclusions}

The mass of the molecular nuclear disk is estimated to be $2 - 6
\times10^7 \msol$ (Oka et al.\ 1998; Pierce-Price et al.\ 2000) with
typical gas temperature $\sim 70$\,K, so the cosmic ray heating in
this region totals $\sim 2 - 6 \times 10^{37}\u erg \ut s -1 $.  The
total cosmic-ray energy losses are five times higher, i.e. $\sim 1 - 3
\times 10^{38}\u erg \ut s -1 $.  Assuming that $\sim 10^{50}\u ergs $
(~10\% of the energy of a typical supernova) goes into particles and
the magnetic field (Duric et al.\ 1995), it leads to one supernova per
$10^4$ years.  Given the high density molecular gas in the nuclear
disk, a SNR lifetime of $\sim2\times10^4$ yr implies a few SNRs in the
nuclear disk, comparable to the number of known SNRs (Gray 1994).  The
uniformity of the dense molecular gas distributed in the nuclear disk
is inferred from the fact that $\sim$60\% of all known remnants in
this region interact with molecular gas versus a value of 10\% in the
Galactic disk (Yusef-Zadeh et al.\ 2007b).  Assuming a Miller-Scalo
IMF, the above estimate of the SN rate implies a star formation rate
($>$ 5 \msol) of $\sim 2.5\ee -3 $ \msol yr$^{-1}$ (e.g., Condon
1992).  Given that $\sim$5\% of the molecular gas in the Galaxy
resides in the nuclear disk, the estimated star formation activity per
unit mass, as traced by SN rate, is two orders of magnitude lower than
that in the Galactic disk.  Thus, star formation is fairly inefficient
in this region of the Galaxy.  Fatuzzo, Adams \& Melia (2006) also
suggest that the increased ionization resulting from the interaction
of supernova remnants with molecular clouds acts to suppress star
formation.

Increased ionization due to enhanced cosmic rays is responsible for
the lower efficiency of star formation in the Galactic nuclear disk
than in the Galactic disk.  Sgr B2 is the best example of current
massive star formation in the nuclear disk, but even there the star
formation per unit mass in Sgr B2 is an order of magnitude lower than
that in W49 and W51, massive star forming regions in the main spiral
arms of the Galaxy (Gordon et al.\ 1993).  All other massive clouds in
the nuclear disk show even lower efficiency of star formation than in
Sgr B2.  For example, the giant molecular cloud G0.25+0.01 has a star
formation efficiency of 0.1\%, roughly 30 times less than that in the
disk of the Galaxy (Lis et al.\ 2001).  The lack of numerous H$_2$O
and methanol masers usually associated with early phases of star
formation, especially in light of the large reservoir of warm and
dense molecular clouds in this region also indicates that the overall
star formation rate in the molecular nuclear disk is generally low.

In conclusion, we have presented the evidence of cosmic rays in Sgr
B2, arguably, the most massive star forming regions in the Galaxy.  We
have also outlined a simple picture of the heating of molecular gas by
cosmic rays and its consequent high ionization fraction can delay the
formation of stars, suppress the formation of low-mass stars in a high
pressure environment.  The impact of the relativistic component of the
ISM with molecular clouds has important implications for the mode of
star formation and also the type of energetic activity found in nuclei
of galaxies.  A recent CO (7--6) line observations of NGC 253
(Bradford et al.\ 2003) also supports a picture in which cosmic rays
are responsible for heating the molecular gas in the nucleus of this
starburst galaxy.
 
Acknowledgments: We thank the referee and the editor, John Scalo, for 
useful comments.

\newcommand\refitem{\bibitem[]{}}

\begin{figure}
  \includegraphics[scale=0.4,angle=0]{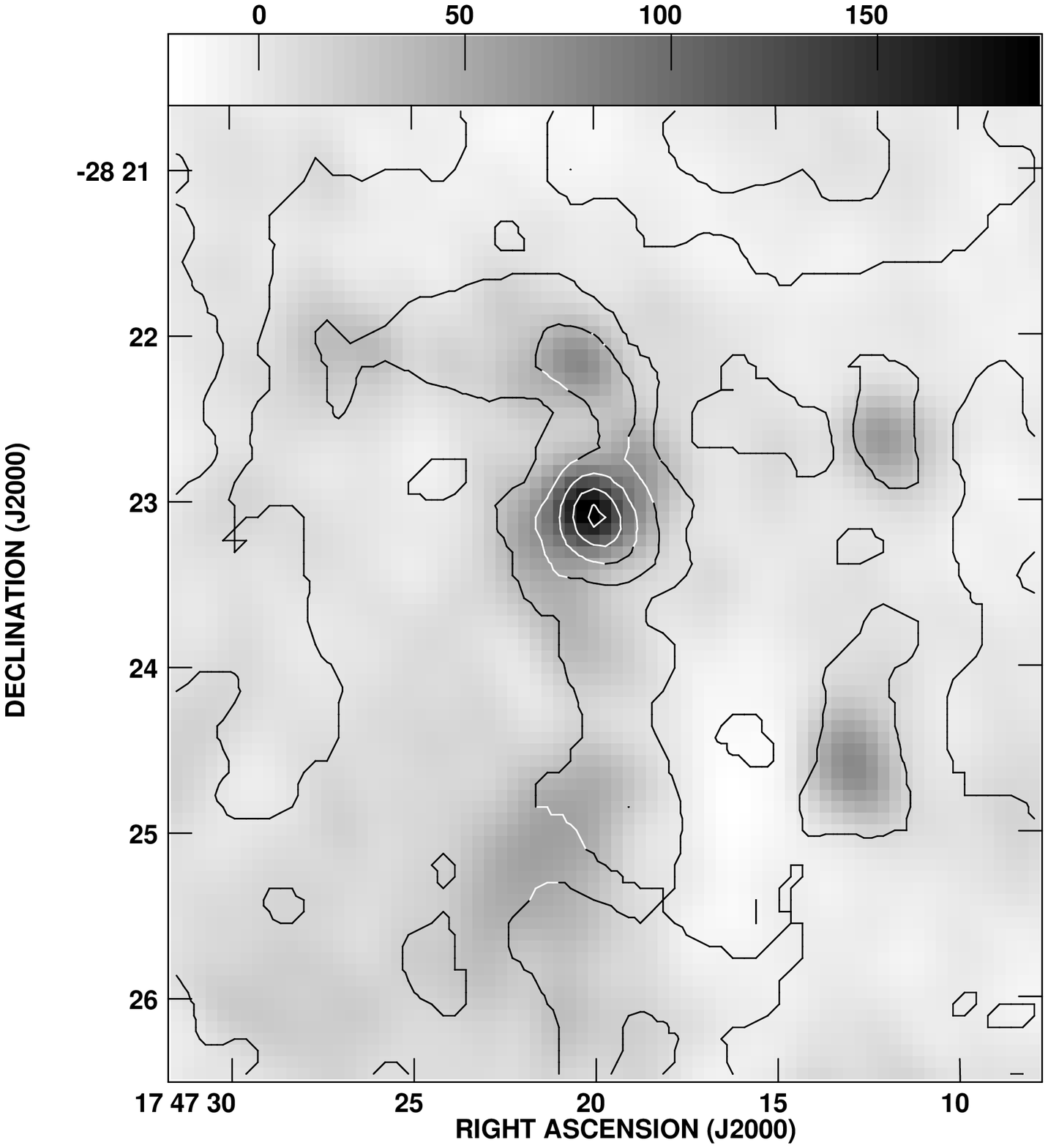}
  \includegraphics[scale=0.6,angle=0]{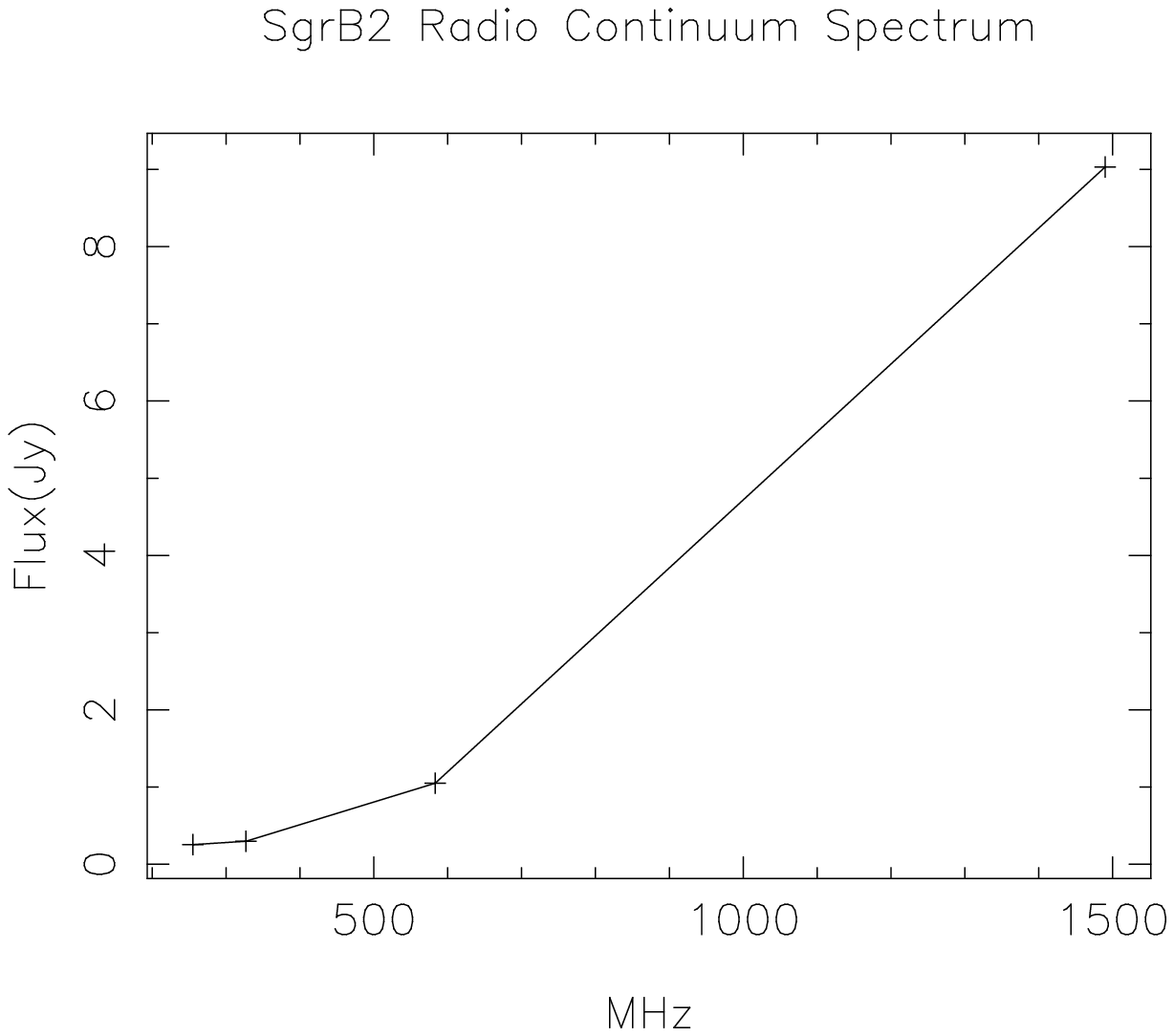}
\caption{{\bf (a) Left} Contours of 255 MHz emission at 30, 50, 70,
90, 120, 160, 200 mJy beam$^{-1}$ with spatial resolution of
22.5\arcs\ and 16.8\arcs\ (PA=19$^0$) are superimposed on a grayscale
image of the F source in Sgr B2 at 327 MHz (Nord et al.\ 2004).  The
peak position coincides with the cluster Sgr B2 F. The grayscale
ranges between -20 and 188 mJy beam$^{-1}$.  The 255 MHz map has used
UV data from 0.04 to 17 k$\lambda$ to bring out extended features.
The RMS noise is about 10 mJy beam$^{-1}$.  The compact sources show a
a positional accuracy 3-4$''$ compared to higher frequency
observations.  {\bf (b) Right} The spectrum of Sgr B2-F at 255, 327,
583 and 1490 MHz, corrected for free-free absorption due to foreground
gas (see text).  Crosses indicate 1$\sigma$ error bars.}
\end{figure}  
\end{document}